 \documentclass[12pt,a4paper]{report}

 \usepackage[T1]{fontenc}
 \usepackage[latin1]{inputenc}
 \usepackage[english]{babel}

 \usepackage{amsmath}
 \usepackage{amsfonts}
 \usepackage{amssymb}
 \usepackage{mathrsfs}
 \usepackage{color}
 \usepackage{ulem}

 \usepackage{fullpage}
 \usepackage[top = 2.5cm, bottom = 2.5cm, right = 2.5cm, left = 2.5cm]{geometry}

 \usepackage{multirow}

 \usepackage{graphicx}

\begin{document}

 \begin{center}
	\Huge{\textbf{Hydrogen Stark broadened Brackett lines\\[1cm]}}
	\Large{C. Stehl\'e and S. Fouquet\\[1cm]}
	\normalsize{LERMA, UMR8112, Observatoire de Paris, CNRS et Universit\'e Pierre et Marie     Curie,\\ 5 place Jules Janssen, 92195 Meudon\\[1cm]}
	{chantal.stehle@obspm.fr}\\
	{sylvain.fouquet@obspm.fr}\\[3cm]
	\large{\textbf{Abstract}}
 \end{center}

	Stark broadened lines of the hydrogen Brackett series are computed for the conditions of stellar atmospheres and circumstellar envelopes. The computation is performed within the Model Microfield Method, which includes  the ion dynamic effects and makes the bridge between the impact limit at low density and the static limit at high density and in the line wings.
The computation gives the area normalized line shape, from the line core up to the static line wings.
\newpage

	\section*{1. Introduction}

	Hydrogen is the most abundant element in the Universe. Its broad lines give noticeable features in the spectra of stellar atmospheres \cite{1, 2, 3, 4}. These lines are very sensitive to the interaction between hydrogen radiating atoms and the surrounding charges, electrons and ions (mostly protons), which is connected to the random electric field generated by these charges. The electric field has two components, with different time scales: the rapidly varying electronic field and the slowly varying ionic electric field. The net field induces a strong mixing of  the atomic states  with the same principal quantum number, from which the Stark broadening originates. Astrophysical applications need to know the full line shape, from line centre up to the line wings, for a wide range of plasma conditions (i.e. electron density $N_e$ and temperature $T$). 
		
	Whereas hydrogen line shapes have been widely used in stellar physics for the determination of gravity and/or temperature in the visible and UV ranges, the recent instrumental developments on various telescopes, such as AMBER \cite{5} and CRIRES \cite{6} on VLTI and VLT, require the availability of precise hydrogen lines in the infrared (i.e. between 1 and 5 $\mu$m). The latter are badly known \cite{7}.

	This paper  thus aims at providing a coherent description of the line shapes of Brackett $\alpha$, $\beta$, $\gamma$, which connects the levels of principal quantum numbers $n$ equal to 4, for the lower state of the transition, and $n'$ equal to 5, 6 and 7, for the upper state, and have central wavelengths of 4.05, 2.63, and 2.12\;$\mu$m. The choosen plasma conditions  are relevant to stellar photospheres and circumstellar environments (electron densities between $10^{10}$ and $10^{19}\,$cm$^{-3}$ and temperatures between $10^3$ and $10^7$\,K). The plasma charges are assumed  to be electrons and protons. This is a standard approximation for this type of study, although some improvements may be possible by including the effect of ionization of He  (which is, by number of atoms, only 10\% as abundant as hydrogen) and traces of heavier elements.	
	
	We  suppose that plasma collective effects are included in the Coulomb interactions between the hydrogen bound electron and the plasma charges  by  Debye screening. This requires  that the Debye length $\lambda_d$,  is larger than the mean distance $r_0$ between the protons, or that the parameter $a = r_0 / \lambda_d$ is smaller than unity. Our tables will thus be limited to the values of temperatures and densities satisfying the condition: 

\begin{equation}
	N_e < 2.6\;10^9\,T^3\quad(\text{cm}^{-3},\;K),
	\label{eq1}
\end{equation}

	which is fulfilled for standard stellar atmospheres.

	We assume also that  the proton/electron density is small enough to ensure that each line 4-n' (n'= 5, 6 or 7) remains distinguishable from the subsequent  line 4-(n'+1) of the Brackett series. Using the Inglis-Teller \cite{8} criterium, the upper limit  to the electron density $N_e$ is thus 

\begin{equation}
	\log(N_{e,max}) = 22 - 7.5\log(n')\quad(cm^{-3}) .
	\label{eq2}
\end{equation}

	This respectively  gives $\log(N_{e,max}) = 16.75, 16.16$ and 15.66 for the $n'$ (= 5, 6 and 7) values  considered in this paper. However, in order to allow interpolation within the tables for astrophysical applications, the line shapes for higher values of $N_e$, up to a decade, have been computed. 
	
	Different methods can be used  to generate high quality spectral line shapes for hydrogen lines perturbed by protons and electrons:   Molecular Dynamics for  describing precisely the ion dynamics effects in the line center \cite{9,10}, quantum theory for the electron contribution to the line wings \cite{11},  short range H-H+ molecular interactions also for the line wings, leading to the apparition of quasi-molecular satellites, which are observed in  the atmospheres of white dwarfs \cite{12}. However, they are limited either to part of the profile, or to restricted plasmas  conditions, or to simple lines like Lyman or Balmer lines. Thus, as for astrophysical purposes, the tabulations go from the line centre up to the line wings, it is necessary to find a compromise between accuracy and  description of the whole profile. 
	
	In this context, the tabulations of Vidal et al.\cite{13}, using Unified Theory for the electrons and static approximation for the ions, have been used for stellar atmospheres, despite the intrinsic lack of accuracy in the line center due to the neglect of ion dynamics effects. The tabulations of  Stehl\'e et al., for the Lyman , Balmer \cite{14, 15} and Paschen lines \cite{16}, using Model Microfield Method, which was initially developed by Brissaud et Frisch \cite{17, 18} brought an important improvement by taking into account the ion dynamics effects.  They are  now used for atmospheres and for the computation of radiative diffusion processes in the radiative stellar envelopes \cite{19}. In the case of partly ionized plasmas, like for the atmospheres of cool stars, the contribution of neutral broadening by hydrogen has to be included in the line shape, especially in the line wings \cite{20, 21}. This effect will be neglected in the following. 

	 In this paper we shall present the Stark broadened profiles of Brackett lines. They will be computed within the formalism of Model Microfield Method, hereafter denoted by MMM. We shall neglect the fine structure effects, which play a role, at low density, in the core of the lines with low $n$ quantum mumbers, like Ly$\alpha$ or H$\alpha$ \cite{22}). 
	 
For Br$\alpha$ , the profiles are computed at  densities log$_{10}$ (N$_e$ (cm$^{-3}) )$  ranging from  10 to 18.5, by step of 0.5. For each density, the profiles are computed at  temperatures equal to 1000, 2500, 5000, 10000, 19550, 39810, 79810, 158500, 316200, 63100 and 1259000 K, assuming that condition  \ref{eq1} is satisfied. For Br$\beta$ , the computation stops at  log$_{10}$ (N$_e$ (cm$^{-3}) )$ = 18 and for Br$\gamma$  at 17.5, as explained previously.  
 
	\section*{2. Method}

	The broadening of spectral lines results from the interactions between the radiating hydrogen atom and the free ions and protons. These two  contributions can be described in terms of interaction potentials, with the corresponding electronic and ionic plasma microfields ${\bf F}_{el}$ and ${\bf F}_{ion}$.  Neglecting quadrupolar and other contributions that  play a role at high densities \cite{23}, the dipolar potential of interaction between the bound electron and the microfields may be written as:

\begin{equation}
	V(t) = -{\bf d} .({\bf F}_{el}(t) + {\bf F}_{ion}(t)) .
	\label{eq3}
\end{equation}

	The spectral line profile $I(\omega)$ (with area normalized to unity) is thus defined in the Liouville space \cite{24,25}, spanned by the states $| i, f\gg$  (which stands for $| n_i,l_i, m_i ; n_f, l_f, m_f\gg$) as

\begin{equation}
	I(\omega) = \frac{1}{  \pi  \sum\limits_{i,f} {\bf d}_{i,f} .{\bf d}_{i,f}^{*}  }
	Re \sum_{i,f,i',f'} {\bf d}_{i,f} . {\bf d}_{i',f'}^{*}  <{\bf U}(\omega)>_{el,ion;\, if;i'f'} ,
	\label{eq4}
\end{equation}

where $<{\bf U}(\omega)>_{el,ion}$ is the Fourier transform of the evolution operator of radiating Hydrogen atom in the Liouville space, averaged over the realizations of the stochastic dynamic electronic and ionic microfields ${\bf F}_{el}$ and ${\bf F}_{ion}$. The term  ${\bf d}_{i,f} . {\bf d}_{i',f'}$ is the product of  dipole operator elements between initial low states (denoted by $i, i'$) and final upper states (denoted by $f, f'$), of the hydrogen bound electron. As fine structure and inelastic effects are neglected, one has  $E_f - E_i = E_f' - E_i' =\hbar  \omega_{0}$.

	The two microfields are stochastic processes. It is thus possible to define two distribution functions $P({\bf F})$ \cite{26,27}, respectively associated to the slowly varying ionic and rapidly varying electronic microfields. In order to take into account the dynamic effects of these microfields, a model for the dynamical statistics of field fluctuations is necessary. In MMM, the microfield fluctuations are handled with a statistical process model, where the microfield (electronic or ionic) is assumed to be constant during a given time interval. The microfield then jumps instantaneously to another constant value for the next time interval. The jumping times are assumed to follow a Poisson law, with a field dependent frequency $\nu(F)$. The jumping frequency $\nu(F)$, is chosen to reproduce the true field autocorrelation function \cite{17}, \cite{18}, \cite{28}.
	This method has been tested against asymptotic impact and quasi-static limits and has been proved to lead to very good results for hydrogen \cite{29} and hydrogenic ion lines \cite{28}. 

	The method has been already described in Stehl\'e and Hutcheon \cite{15}, and we refer the reader to this paper for the details. An important point is that it is possible to disentangle the contributions of ions and fast electrons by introducing a frequency dependent electronic relaxation operator $\gamma_{el}(\omega)$, which is independent from ionic fields, and thus may be computed separately. The Fourier transform of the evolution operator, $<{\bf U}(\omega)>_{el}$, averaged over all the realisations of the electronic fields,   may be written as 

\begin{equation}
	<{\bf U}(\omega)>_{el} = i \, [\Delta \omega  {\bf I}+ i  \gamma_{el}(\omega)]^{-1} \,  , 
	\label{eq5}
\end{equation}

where  $\Delta \omega = \omega - \omega_{0}$ is the detuning from line center and 
${\bf I}$ is the identity operator in the Liouville space.

	Thus,  this electron damping is first computed to account for  average effect of the electronic fields. Then, the static Fourier transform of the evolution operator, $<{\bf U}(\omega)>_{el,ion}$,  averaged over the realisations of the electronic and ionic fields, may be written as :

\begin{equation}
	<{\bf U}(\omega)>_{el,ion} = i \, \int {P({\bf F}_{ion}) d{\bf F}_{ion} } \,
	[\Delta \omega {\bf I}  + {\bf d}. {\bf F}_{ion} + \gamma_{el}(\omega) ]^{-1} \, . 
	\label{eq6}
\end{equation}

	The MMM expression is more complex (\cite{17,18}) than this one, which corresponds to the usual Unified Theory with static ionic fields. 
	
	In order to reduce the dimensions of the Liouville space, (16 $\times$ 25 = 400 states for Br$\alpha$, for instance), we use the formalism of the reduced Liouville space, which takes advantage of the invariance of the different operators,   like \textbf{d.d},  by angular average over all the orientations of the electric fields and of the fact that the dipole tensor \textbf{d} in equation \ref{eq4} is of rank 1 (see for instance \cite{28, 29}). However, the number of reduced states $| n_i, l_i; n_f, l_f \gg$ (with $| l_i - l_f | \leq 1$) remains important (i.e. 10 reduced Liouville states of rank 1 for Br$\alpha$). Thus, we use another approximation, already called « isotropic » approximation in \cite{15}, which uses the diagonal form $\gamma_{el}^{iso}(\omega)$, with equal diagonal matrix elements, instead of $\gamma_{el}(\omega)$. This scalar tensor  is deduced from  the pure electronic profile $I(\omega)_{el}$ by  the relation,

\begin{equation}
	I(\omega)_{el} = -\frac{1}{\pi} \, Im [ \Delta \omega + i \gamma_{el}^{iso}(\omega)]^{-1} \, , 
	\label{eq7}
\end{equation}

or 

\begin{equation}
	\gamma_{el}^{iso}(\omega) =  \frac{\sum\limits_{i, f; i', f'} {\bf d}_{i, f} . {\bf d}_{i', f'}^{*} \gamma_{el; \, if; i'f'}(\omega)} {\sum\limits_{i, f}  {\bf d}_{i, f} . {\bf d}_{i, f}^{*}} \, .
	\label{eq8}
\end{equation}

	The Stark profile is thus obtained after computing the average over electronic field values, which gives $\gamma_{el}^{iso}(\omega)$, and then the average over the ionic fields, both using the formalism of Model Microfield Method. The third step is then the Doppler convolution, which gives the final line shape.

	\section*{3. Results}

	The three Brackett  $\alpha, \beta, \gamma$ line shapes have been studied in a wide range of stellar conditions, but we report hereafter only a selection of results, relative to the line center and the line wings.

		\subsection*{3.1 Line center}

\begin{figure}[h]
	\begin{center}
		\includegraphics[width=0.7\linewidth]{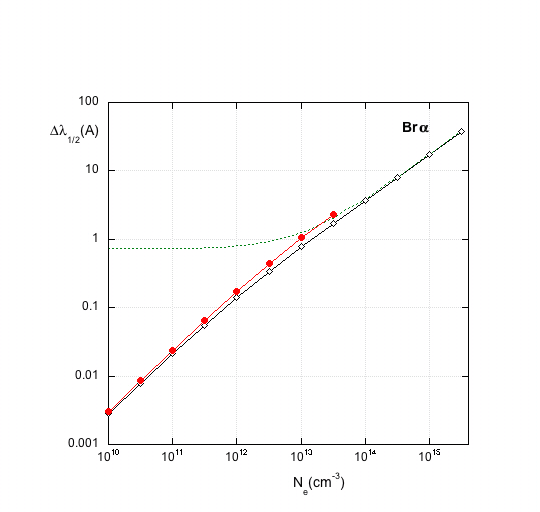}
		\caption{\itshape HWHM of Br$\alpha$ at $2500\,$K, in Å, versus electron density $N_e$ in cm$^{-3}$ (black and squares: MMM Stark only;  red and circles: analytical impact formula; dashed green: MMM Stark profile with Doppler convolution).}
		\label{fiq1}
	\end{center}
 \end{figure}

\begin{figure}[h]
	\begin{center}
		\includegraphics[width=0.7\linewidth]{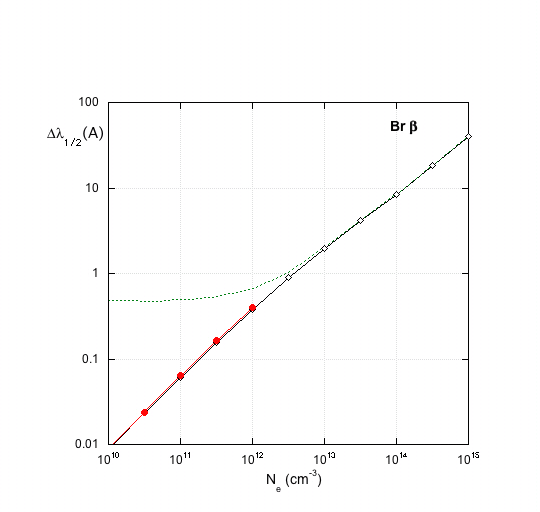}
		\caption{\itshape Same as figure \ref{fiq1} for Br$\beta$.}
		\label{fiq2}
	\end{center}
 \end{figure}

\begin{figure}[!h]
	\begin{center}
		\includegraphics[width=0.7\linewidth]{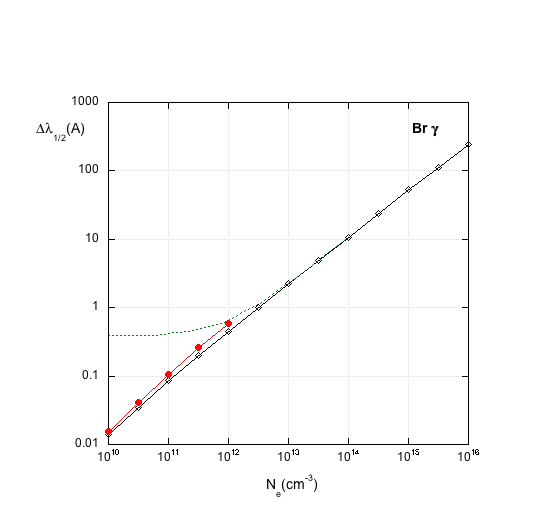}
		\caption{\itshape Same as figure \ref{fiq1} for Br$\gamma$.}
		\label{fiq3}
	\end{center}
 \end{figure}

	We present in figures \ref{fiq1}, \ref{fiq2} and \ref{fiq3} the variations of the half-width (HWHM) of Br$\alpha$,  Br$\beta$  and Br$\gamma$ lines for different values of the electron density. This  quantity is the detuning from the line center, at the point where the profile reaches half the maximum value of the line profile (which is not necessarily at the line center, as will be discussed below). The figures show the half width of the MMM profile with and without Doppler effect and also the value of the impact half-width in its own validity range. 
	
	Hence, impact limit has been proved to be valid, for both electron and ion contributions, in the line center and at low densities. Moreover, it has been proved that the value of the impact width is analytical and that the corresponding profile is Lorentzian in the line center \cite{30}. The validity condition of the impact limit is that the half-width value is smaller than the ion plasma frequency. Let us take the example of of Br$\alpha$.  At 10$^{12}\,$cm$^{-3}$ and 2500\,K the ion plasma frequency is equal to 2.6 10$^9$ rd.s$^{-1}$ ,whereas the ion impact contribution to the HWHM is equal to 1.8 10$^9$ rd.s$^{-1}$. At 10$^{14}$ cm$^{-3}$, they are respectively equal to 2.6 10$^{10}$ and 3.9 10$^{10}$. The impact limit should thus be reached gradually as the density decreases below 10$^{13}$ cm$^{-3}$. 
	
The figure \ref{fiq1}, relative to Br$\alpha$, shows indeed that the half-width  of the  Stark profile (black), tends to converge towards   the impact analytical  limit (red) at these low densities. However, the convolution with the Doppler profile increases  the half-width value. As a consequence, the half width is dominated by the Doppler  broadening at low densities.   Similar behaviour occurs for the other lines, as may be  on figures \ref{fiq2} and \ref{fiq3} for Br$\beta$ and Br$\gamma$.

\begin{figure}[!h]
	\begin{center}
		\includegraphics[width=0.6\linewidth]{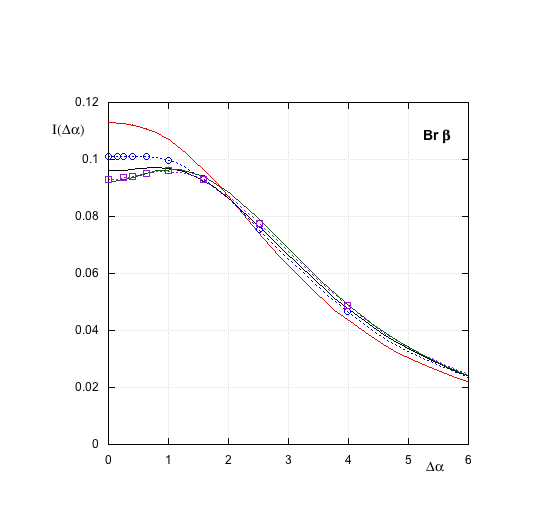}
		\caption{\itshape MMM  Br$\beta$ line at 10$^{15}\,$cm$^{-3}$, versus $\Delta \alpha = \Delta \lambda / F_0$ without Doppler convolution for different temperatures (red : 1000\,K; blue with empty circles : 2500\,K; violet with empty squares: 10000\,K;  green: 19950\,K).}
		\label{fiq4}
	\end{center}
 \end{figure}

At higher densities,  the ion dynamic effects become smaller in the line center (they are negligible in the line wings,  as will be seen below), and the line shape departs from the Lorentzian shape. We found that, depending on the temperature conditions, the Brackett $\beta$ line may present a small dip in the line center at moderate densities of 10$^{15}\,$cm$^{-3}$ and for the largest temperatures as seen in figure \ref{fiq4}. This effect was well known in earlier tabulations for the Lyman$\beta$ and H$\beta$ lines. The dependence versus the temperature is a consequence of the the electron impact broadening, which varies in T$^{-1/2}$ , and which may fill (at low T values) or not (at large T values) the central dip.

		\subsection*{3.1 Line shapes}

	Another typical behaviour of hydrogen lines is the convergence towards the (Holtsmark) static limit in the line wings, which scales as $|\Delta \omega|^{-5.2}$ when the line shape is expressed in angular frequency units (which is the appropriate unit for the line shape computations). However, traditionally, the line intensity was expressed in units of $\Delta \alpha = \Delta \lambda / F_0$ , where $F_0$ (esu) = 1.25 10$^{9}$ $(N_e)^{2/3}$ is the normal Holtsmark field. In these units, the Holtsmark limit is given by :

\begin{equation}
	I(\Delta \alpha ) = \frac{K_{\alpha} }{| \Delta \alpha | ^{5/2} }	
	\left(   \frac{  \lambda_0/F_0 + \Delta \alpha }{ \lambda_0/F_0 }   \right)^{1/2}
	\simeq  \frac{K_{\alpha} }{| \Delta \alpha | ^{5/2} } 	\,\,\,
\rm{for} \,\,\,
       |\Delta \alpha| << \lambda_0/F_0
	\label{eq9}
\end{equation}

	where $\lambda_0$ is the central wavelength, and $K_{\alpha}$ a constant, which depends on the transition,

\begin{equation}
	K_{\alpha} =  \text{ 1.512  for Br$\alpha$, 2.401 for Br$\beta$ and 2.926 for Br$\gamma$}
	\label{eq10}
\end{equation}

\begin{figure}[h!]
	\begin{center}
		\includegraphics[width=0.9\linewidth]{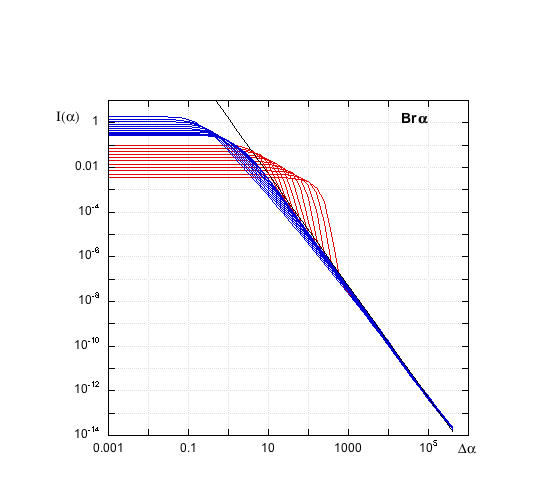}
		\caption{\itshape MMM Br$\alpha$ line at N$_e$=10$^{12}$ cm$^{-3}$, with (red) and without (blue) Doppler broadening, and the asymptotic Holstmark limit (black)  in $|\Delta \alpha|^{-5/2}$ for 11 different temperatures  in K (1000, 2500, 5000, 10$^4$, 2 10$^4$, 3.98 10$^4$, 7.94 10$^4$, 1.58 10$^5$, 3.16 10$^5$, 6.31 10$^5$, 1.26 10$^6$). The corresponding values of the profiles in the line center are respectively : 		
		0.28 ; 0.30 ; 0.34 ; 0.46 ; 0.56 ; 0.69 ; 0.88 ; 1.23 ; 1.45 ; 1.87, without Doppler effect (blue), 
		 and : 
		0.102 ; 0.071; 0.053; 0.039; 0.028 ; 0.020; 0.014; 0.010 ; 0.0072 ; 0.0051; 0.0036
		including Doppler  effect (red).}
		\label{fiq5}
	\end{center}
 \end{figure}

\begin{figure}[h!]
	\begin{center}
		\includegraphics[width=0.9\linewidth]{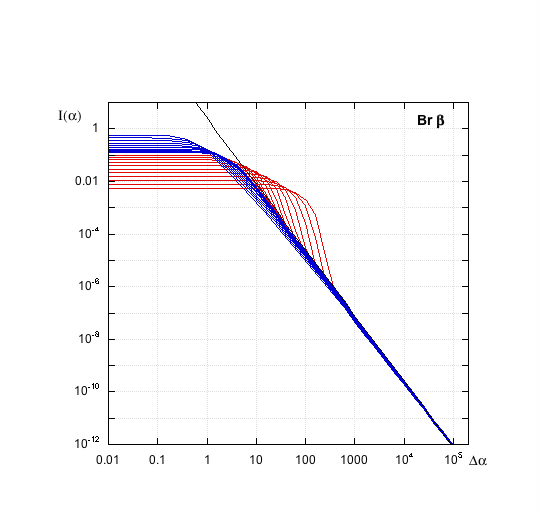}
		\caption{\itshape Same as Fig.\ref{fiq5} but for Br$\beta$, 
		The values of the profiles in the line center are, for the 11 different temperatures of  Fig.\ref{fiq5},  respectively : 	
		0.124 ; 0.129 ; 0.136 ;  0.149 ;  0.168 ;  0.195 ;  0.233;  0.285 ;  0.355 ;  0.451 ;  0.579 
without Doppler effect (blue),  
and
  0.099;  0.082 ;  0.067 ;  0.053 ;  0.040 ;  0.030 ;  0.0215 ;  0.015 ;  0.011 ;  0.0078; 0.0056
		including Doppler  effect (red).}
		\label{fiq6}
	\end{center}
 \end{figure}

\begin{figure}[h!]
	\begin{center}
		\includegraphics[width=0.9\linewidth]{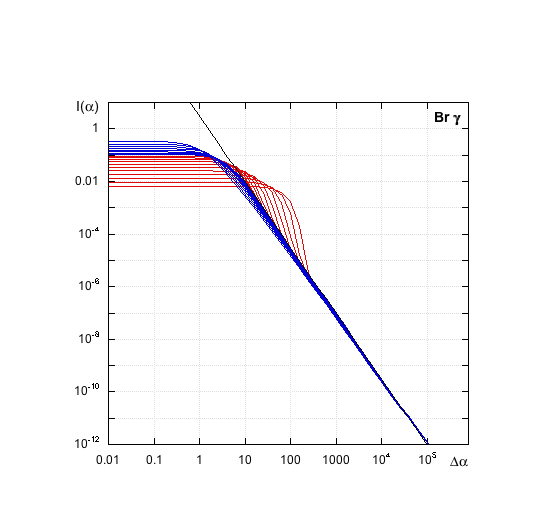}
		\caption{\itshape Same as Fig.\ref{fiq5} but for Br$\gamma$, 
		The  values of the profiles in the line center are, f for the 11 different temperatures of  Fig.\ref{fiq5},  respectively : 	
		0.107 ;  0.107 ;  0.109 ;  0.113 ;  0.121 ;  0.132 ;  0.150 ;  0.175 ;  0.211 ;  0.260 ;  0.327 
without Doppler effect (blue),  
and
  0.091 ;  0.080 ;  0.069 ;  0.057 ; 0.0450 ;  0.034 ;  0.025 ;  0.018 ;  0.013 ;  0.0094 ;  0.0067
		including Doppler  effect (red).}		
		\label{fiq7}
	\end{center}
 \end{figure}

	This variation in $\Delta \alpha$ of eq. \ref{eq9} introduces, at large detunings, a "trivial" asymmetry between $I(\Delta \alpha)$ and $I(-\Delta \alpha)$ (which does not exist between $I(\Delta \omega)$ and $I(-\Delta \omega)$). The figures \ref{fiq5}-\ref{fiq7} show the profiles $I(\Delta \alpha)$ of Br$\alpha$, Br$\beta$ and Br$\gamma$ lines at 10$^{12}\,$cm$^{-3}$ and various temperatures. The profiles, including Doppler effects are reported in red color, and the pure Stark profiles in blue. 
	
	The profiles are area normalized, i.e. :

\begin{equation}
	\int_{-\infty}^{+\infty} \, I(\Delta \alpha) \times d(\Delta \alpha)   = 1
	\label{eq11}
\end{equation}

	As a consequence, broad profiles have small values of $I(\Delta  \alpha = 0)$. For the density considered in the figures \ref{fiq5}-\ref{fiq7}, the Doppler profile dominates the Stark profile in the line center and becomes indistinguishable, as expected, in the line wings, where they follow the asymptotic limit of eq. \ref{eq9}. At higher densities (not reported here), the Doppler width is smaller than the Stark width and Doppler convolution is no longer necessary. 

	\section*{\textbf{4. Conclusions}}

	This study shows that Stark broadened infrared Brackett lines of hydrogen follow the same trends as the lines of lower series which are more known theoretically and experimentally. This study will allow missing absorption in the spectra of stellar atmospheres in the infrared due to the lack of data to be filled in a next future. Dedicated experiments and comparisons with other methods, for instance FFM \cite{32}, would be helpful to test these theoretical results, which will be also constrained by observational data. The corresponding tables will be accessible at http://amrel.obspm.fr/stark-h.

	\section*{\textbf{5. Acknowlegdments}}

	We would like to thank Pauline B\'eghin, Edrice Bouteldja, Aur\'elie de Paz and C\'ecile Turc, students at the University of Pierre et Marie Curie for their contribution in checking the data.

\end{document}